\documentclass[fleqn,10pt,table]{wlscirep}
\usepackage{amsmath,nccmath}
\usepackage{graphicx}
\usepackage{dcolumn}
\usepackage{bm}
\usepackage[utf8]{inputenc}
\usepackage[T1]{fontenc}
\usepackage{mathptmx}
\usepackage{duckuments} 
\usepackage{pgfplots}
\usepgfplotslibrary{external}
\usetikzlibrary{shapes,arrows,calc,fit,arrows.meta,external,positioning}
\pgfplotsset{compat=1.16}
\title{Laser excited super resolution thermal imaging for nondestructive inspection of internal defects}

\author[1,*]{Samim Ahmadi}
\author[1]{Julien Lecompagnon}
\author[1]{Philipp Daniel Hirsch}
\author[2]{Peter Burgholzer}
\author[3]{Peter Jung}
\author[3]{Giuseppe Caire}
\author[1]{Mathias Ziegler}
\affil[1]{Bundesanstalt für Materialforschung und -prüfung (BAM), 12200 Berlin, Germany}
\affil[2]{Research Center for Non Destructive Testing, 4040 Linz, Austria}
\affil[3]{TU Berlin, Communications and Information Theory, 10587 Berlin, Germany}
\affil[*]{samim.ahmadi@bam.de}

\begin{abstract}
A photothermal super resolution technique is proposed for an improved inspection of internal defects. To evaluate the potential of the laser-based thermographic technique, an additively manufactured stainless steel specimen with closely spaced internal cavities is used. Four different experimental configurations in transmission, reflection, stepwise and continuous scanning are investigated. The applied image post-processing method is based on compressed sensing and makes use of the block sparsity from multiple measurement events. This concerted approach of experimental measurement strategy and numerical optimization enables the resolution of internal defects and outperforms conventional thermographic inspection techniques.
\end{abstract}
\begin{document}

\flushbottom
\maketitle
%
%
\thispagestyle{empty}

\section*{Introduction}

The nondestructive testing (NDT) of internal defects such as blowholes, inclusions or delaminations is of huge interest in industry. There are several ways to detect internal defects without destroying the specimen such as ultrasonic testing (UT) or radiographic testing (RT). UT is typically not contact-free and suffers from reconstruction accuracy if defects are not oriented perpendicular to the coupled ultrasound. In contrast, RT methods like computed tomography provide reliable and accurate results \cite{dierolf} and are contact-free, but end up to be costly, slow, complex and only suitable ex-situ. Unlike UT and RT, active thermographic testing (TT) represents a contactless, simple, less expensive and in-situ suitable alternative by measuring the infrared (IR) radiation intensity of the specimen with IR cameras \cite{hung}. 

In active TT, light sources such as lasers can be used to generate heat in the specimen \cite{bouzin2019photo,thielapl}. Compared to other light sources such as flash lamps, halogen lamps or LED, lasers do not exhibit spectral overlap with the IR camera \cite{ziegler2018lock} and can be tightly focused which helps to realize structured illumination (SI). Beside light sources, other energy sources, e.g. induction coils and ultrasonic transducers are possible as well for thermal NDT \cite{vavilov2009infrared}. 

The diffuse nature of heat propagation in the material causes a degradation in spatial resolution and therefore also in reconstruction accuracy \cite{APL}. To solve this problem, various measurement and thermal image processing strategies were applied such as pulsed-phase thermography \cite{maldague2002advances} or lock-in thermography \cite{Wallbrink}, making use of the relative amplitude or phase change to a reference area. A relatively new method to circumvent spatial heat blurring is the introduction of virtual waves, which increases the signal-to-noise ratio (SNR) in the measured thermal images by transforming diffuse thermal waves into virtual propagating waves \cite{burg_vw,gm,thummerer2020photothermal}.

Apart from that, so-called optical super resolution (SR) imaging - serving as an alternative measurement strategy to enhance the spatial resolution - gained attention in fields of structured illumination microscopy \cite{muller2016open,heintzmann2017super}.
These SR techniques rely on multiple measurements with a small position shift. The result is a spatial frequency mixing of the illuminated target pattern and the illumination pattern enabling an improvement in spatial resolution. While these optical SR techniques aim to enhance the optical diffraction limit of the imaging system, geometrical SR techniques aim to enhance the resolution of the digital imaging sensors. While the latter approach is known \cite{panagiotopoulou2008super,sakagami2009nondestructive} and already implemented in commercial IR camera systems, a method to overcome the diffusion limit of TT by means of an optical SR analogue was out of reach so far.

Compressed sensing (CS) based algorithms can be used in post-processing which benefit from multiple measurements all referring to a reconstruction result that is sparse \cite{zhu2012faster}. Since defects are sparse in space and CS algorithms rely on reconstructing a sparse data set from given measurements, CS is highly attractive and applicable in NDT scenarios as well. Thus, CS based algorithms based on sparsity regularization with $\ell_1$-minimization or Orthogonal Matching Pursuit (OMP) have been successfully employed to thermographic data improving the thermal image quality \cite{gao2015unsupervised,tang2017smart}. However, the application of a simple $\ell_1$-minimization or OMP would not benefit from multiple measurements as generated by the proposed SI experiments. In contrast, Block CS would be more suitable \cite{eldar2009block} so that algorithms such as Block-FISTA \cite{murray2017super} or Block-OMP \cite{eldar2010block} could profit from many measurements exhibiting a so-called joint or block sparsity domain. 

Novel photothermal SR approaches combine the SI measurement strategy and CS based processing algorithms, such as the iterative joint sparsity algorithm (IJOSP) to better separate two closely spaced defects \cite{APL,QIRT,paper1,ahmadi2}. So far, the suitability of these techniques has been investigated in transmission configuration with anomalies on the backside of the specimen. This paper focuses on the applicability of laser excited super resolution thermal imaging to an additively manufactured stainless steel sample with internal defects as test specimen. 

The major contributions of this paper are listed as follows:
\begin{itemize}
    \item Different SI experiments (laser step and continuous scan in reflection and transmission configuration) are shown to create suitable data to perform SR and to resolve all internal defects (so far in literature: SR suitable experiments only in transmission configuration examining sample surface anomalies).
    \item The thermal image processing and the application of the model-based IJOSP algorithm is described in details and adopted for the investigation of a specimen with internal defects in transmission and reflection configuration (so far in literature: photothermal SR image processing description only available based on transmission setups analyzing sample surface anomalies).
    \item The SR reconstruction results are compared qualitatively as well as quantitatively with conventional laser thermography reconstruction results based on homogeneous illumination.
    The comparison shows that the proposed SR techniques outperform conventional photothermal techniques resolving internal defects in steel with an at least four times better spatial resolution. 
    \item It is discussed which of the proposed SR techniques - based on different experimental setups - exhibits the most promising results in improving the spatial resolution for TT of metals with internal defects.
\end{itemize}

\section*{Methods}

\subsection*{Experimental setup} \begin{figure*}[!b]
\def\svgwidth{\linewidth}
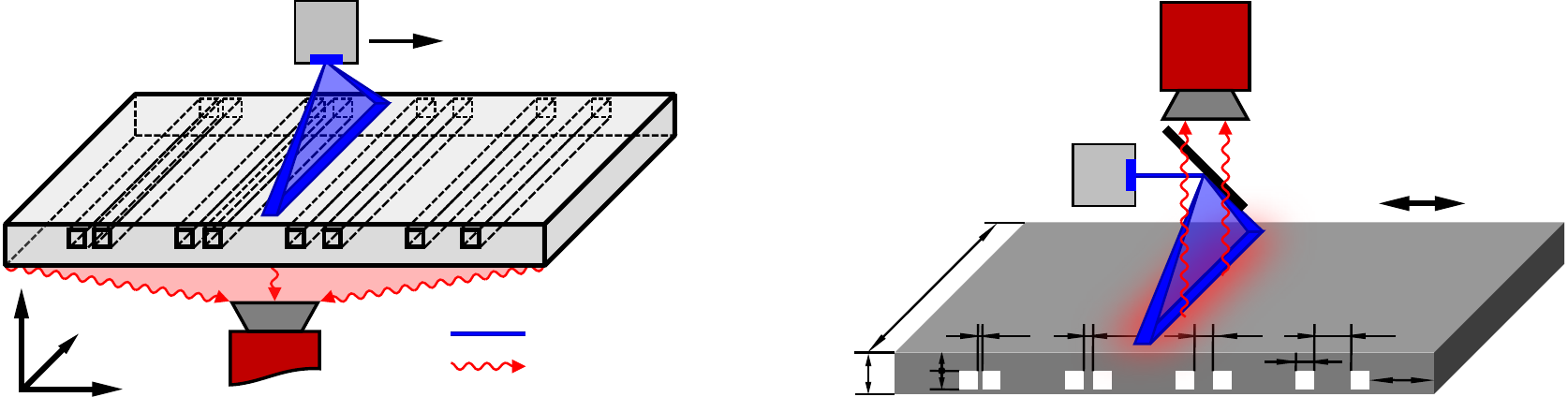 

\caption{\label{exp_setup} The IR camera and the specimen were placed for both configurations on a linear table so that both components are moved with a certain velocity $v_{\mathrm{scan}}$. a) Transmission configuration: The additively manufactured stainless steel specimen (sandblasted from both sides to increase the emissivity) is shown with details inside.
b) Schematic of the setup in reflection configuration. The specimen with eight cavities and different distances is shown (a\,=\,$0.5$\,mm, b\,=\,$40$\,mm, c\,=\,$4.5$\,mm, d\,=\,$7$\,mm).
The arrows in a) and b) represent the direction of the motion.}
\vspace*{\floatsep}
\centering
\includegraphics[width = 0.7 \textwidth, trim = {0cm 0cm 0cm 0cm},clip]{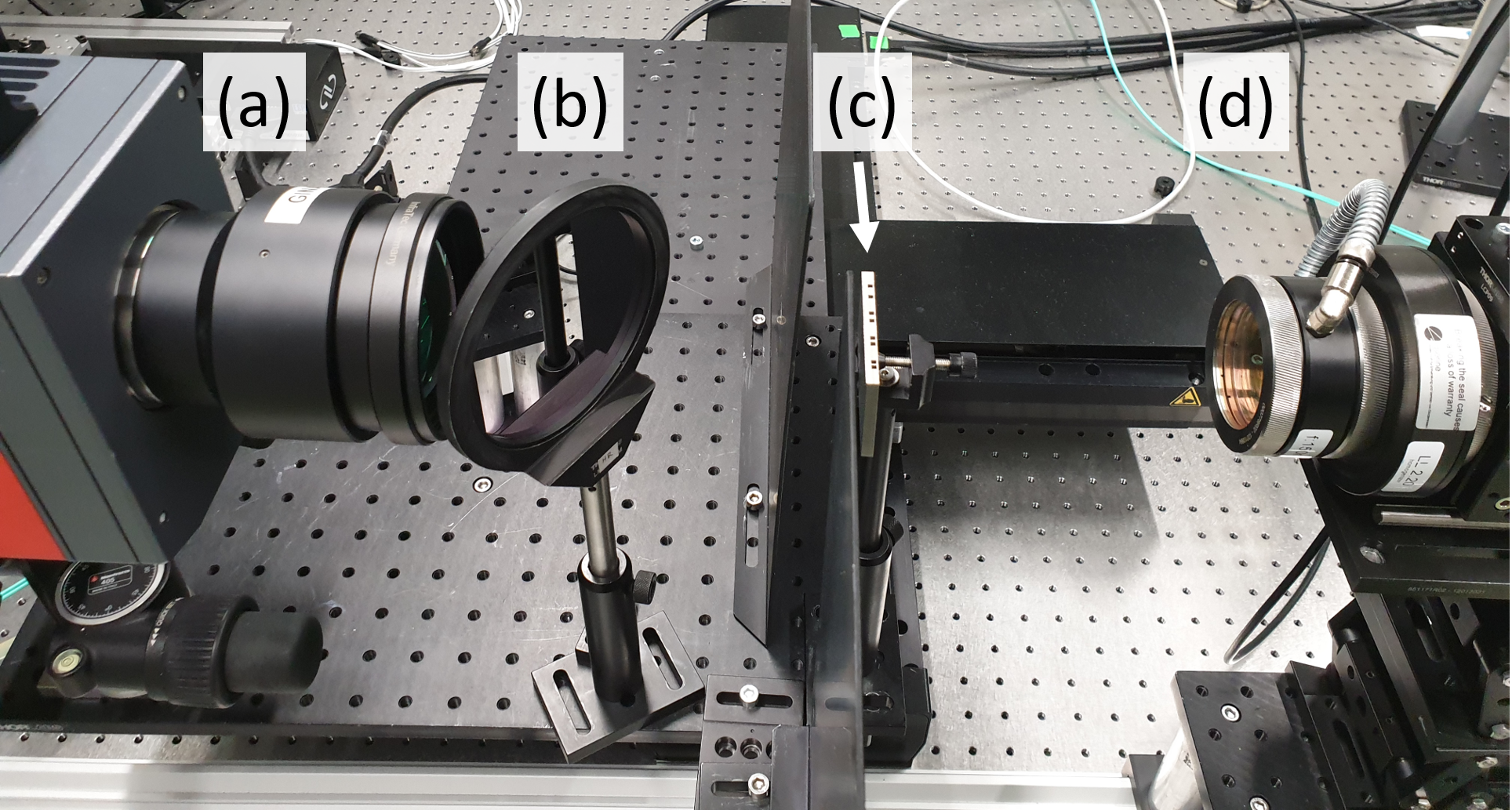}
\caption{\label{real_exp_setup} Exemplary transmission configuration for laser excited super resolution thermography from our laser laboratory at BAM: (a) IR camera, (b) dichroic mirror (used to protect the IR camera from laser beam fractions), (c) investigated stainless steel specimen shown in Figure \ref{exp_setup}, (d) high-power fiber-coupled laser. (a), (b) and (c) are mounted on a linear stage using an optical breadboard. The position of the linear stage is shifted by a motion controller in submillimeter range which is necessary for optical super resolution.}
\vspace{-2mm}
\end{figure*}
\noindent In the following we focus on the measured data obtained by using the transmission and reflection setup shown in Fig. \ref{exp_setup} a) and b), both performed in step scan and continuous scan measurements, respectively. An exemplary experimental thermographic setup is shown as a photograph in Figure \ref{real_exp_setup} which would correspond to the shown configuration in Figure \ref{exp_setup} a). 

As shown in our previous work \cite{ahmadi2}, step scan means that we use laser pulses ($t_{\mathrm{pulse}}$\,=\,$500$\,ms) to heat up the sample, wait until the sample is cooled down ($t_{\mathrm{cooling}}$\,=\,$20$\,s), shift the position slightly with a distance of $\Delta r$\,=\,$0.2$\,mm and repeat for around $250$ individual measurements. In all our measurements we have used a fiber-coupled high-power diode laser with a maximum output power of $530$\,W, a linear shaped spot ($0.4\,\text{mm} \times 17$\,mm) and a wavelength of $940 \pm 10$\,nm. Furthermore, we have used a mid-wave IR camera (Infratec IR9300, $f_{\mathrm{cam}}$\,=\,$100$\,Hz, full frame: $1280 \times 1024$\,pixel, sensitive in $3$-$5$\,$\mu$m wavelength range, NETD - noise equivalent temperature difference of $\sim 30$ mK) which was triggered by a photodiode that recognizes when the laser is switched on. We have measured around $1000$ frames from the beginning of the pulse for each measurement with a pixel resolution of $\Delta r_{\mathrm{cam}} = 54$\,$\mu$m/pixel. This leads to a huge amount of data. In contrast, the continuous scanning method provides a smaller measured data set, since the specimen is scanned continuously while the IR camera is measuring without any intermediate cooling process. Thus, the size of the measured data is controlled by the chosen scanning velocity. The power of the laser has been adjusted in each measurement configuration so that we reach temperature differences of around $\Delta T$\,=\,$3 - 5$\,K for an assumed emissivity of $\epsilon = 1$ (since only relative changes in temperature are of interest, this parameter was not considered critical).

\subsection*{Mathematical model for super resolution laser thermography}
Without restricting the generality of our approach, we simplify the 3dim problem to a 2dim one by using a set of linear defects and a linear laser. We calculate the mean over the vertically arranged pixels (see dimension $y$ in Fig. \ref{exp_setup}) and end up with a problem formulation in the $r$-$z$-domain. In addition, the mean over 315 pixels provides a better SNR of $\Delta T / \text{NETD}\cdot\sqrt{315}\approx 4\, \text{K} / 30\, \text{mK}\cdot 18 = 2400$.

\begin{figure*}[!b] \centering
\includegraphics[width = 0.7\textwidth,trim={50cm 0cm 0cm 0cm},clip]{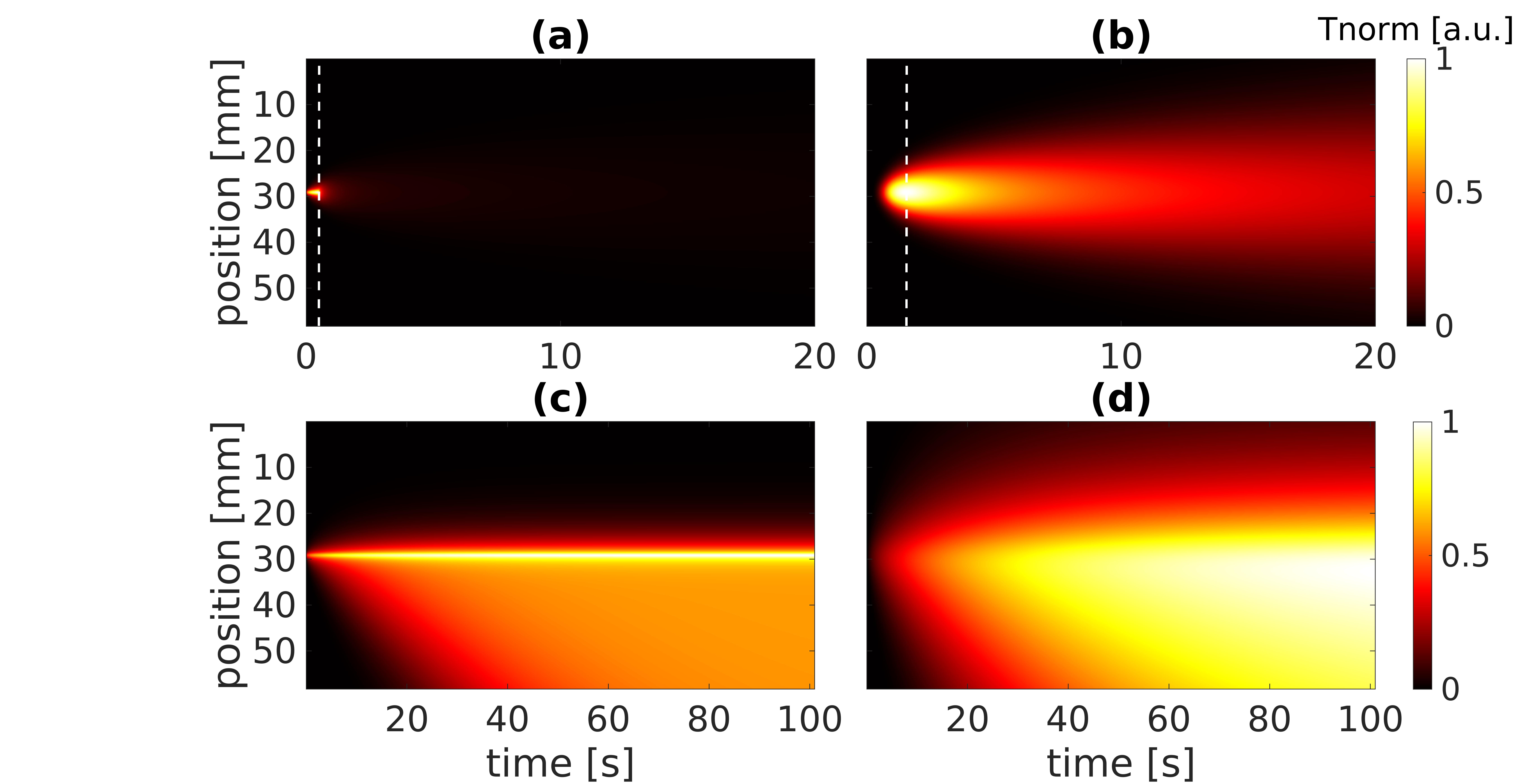}
\caption{\label{psfs} Diagrams of $\Phi$ for a) step scanning, reflection configuration, b) step scanning, transmission configuration, c) continuous scanning, reflection configuration, d) continuous scanning, transmission configuration [$v_{\mathrm{scan}}$\,=\,$1$\,mm/s in c) and d)]. Dashed vertical lines indicate the position of the maximum thermograms used for time dimension reduction.}
\end{figure*}

To describe the measured temperature difference values $T_{\mathrm{meas}} = \Delta T$, we make use of a Green's function with rectangular coordinates as a solution for the heat diffusion equation for a line source and a plate with a finite thickness. In our previous studies \cite{paper1,ahmadi2}, we investigated steel samples which have been blackened on the front side so that the measured temperature from the backside could be described for each measurement $i$\,=\,$1 \dots N_{\mathrm{meas}}$ by
\begin{ceqn}
\begin{align}  \label{main_eq}
    T_{\mathrm{meas}}^i = \Phi *_{r,t} x^i, 
\end{align} \end{ceqn} whereby $\Phi$ represents the heat diffusion followed from illumination and $x^i = I^i \circ a$ represents the element-wise (Hadamard) product of the illumination $I^i$ and absorption pattern $a$ which represents the defects in space \cite{APL,paper1,ahmadi2}. In contrast to our previous model, there is no analytical solution for the case of internal defects to be investigated now, i.e. the $\Phi$ required now is unknown. Since we are not interested in an exact reconstruction of the internal geometry, but only in the best possible separation of closely adjacent internal defects, we can continue to use this approach as an approximation. This means, we pretend to have an exactly describable sample without defects by $\Phi$ and allow the internal defects sparsely distributed in the $r$-$z$-domain to lead to a sparsely distributed contrast in the result $T_{\mathrm{meas}}$, which we interpret as defect position in the $r$-domain. Thus, $\Phi$ can be described by \cite{PSF,ahmadi2}
\begin{ceqn}\begin{equation} \label{phi_eq}
    \Phi (r, z, t) = \frac{2}{4 \pi \alpha \rho c_p}\int_0^t I_t(t-\tilde{t})e^{-\frac{r^2}{4 \alpha \tilde{t}}} \cdot \sum_{n = -\infty}^{\infty} R^{2(n-1)}e^{-\frac{(2nL+z)^2}{4 \alpha  \tilde{t}}} \frac{d \tilde{t}}{\tilde{t}}
\end{equation} \end{ceqn}
for step scanning $\Phi_{\mathrm{step}} = \Phi$, whereby we set $z$\,=\,$0$ for reflection and $z$\,=\,$L$ for transmission configuration, $r$ stands for the position of the horizontally arranged pixels and $t$ stands for the time.
Thus, equation (\ref{phi_eq}) describes a thermal point spread function that considers the convolution with the laser pulse length in step scanning by $I_t$ (convolution in time with the variable $t$). For continuous scanning $\Phi_{\mathrm{cont}}$ we have to substitute $I_t$\,:=\,$1$ since the laser is switched on continuously, $r$\,:=\,$r-v \cdot t$ due to motion consideration and to add an integral over the previous time stamps \cite{ahmadi2,movinglineheating} instead of having an integral due to the pulse length consideration as shown in equation (\ref{phi_eq}). $\rho$ stands for the mass density, $c_p$ for the specific heat, $\alpha$ for the thermal diffusivity, $R$ for the thermal reflectance from the material to air, and $L$ for the thickness of the specimen. We have identified the following values for the material parameters of our investigated additively manufactured stainless steel 316L 1.4404 sample: $\alpha$\,=\,$4\cdot 10 ^{-6}$\,m$^2$/s, $\rho$\,=\,$7990$\,kg/m$^3$, $c_p$\,=\,$500$\,J/kg/K and $R$\,=\,$1$. Indeed, $R$\,=\,$1$ was chosen for the sake of simplicity, but it is a very good approximation since R actually should be around $0.95$. It should be noted that the laser line width is not taken into account in $\Phi$, which means that this quantity has to be considered in $x$ \cite{ahmadi2}.

\begin{figure*}[t]
\centering
\includegraphics[clip, trim={14mm 0 0 0}]{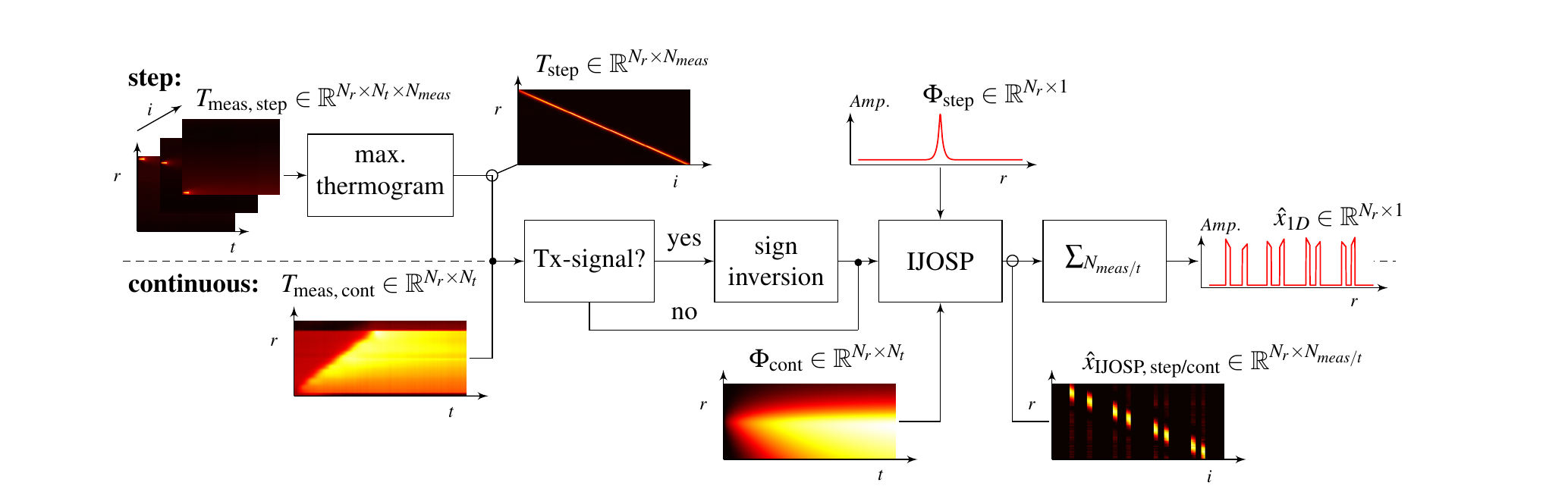}
\caption{\label{flow chart} Flow chart depicting the data analysis procedure discussed in this work. The chart is split in two halves by a dashed line: the bottom half describes the necessary steps for continuous scanning data, while the top half shows the steps for data generated by step scanning. The direction of data flow is from left to right. The inspection of intermediate results is marked by empty circles. Tx = Transmission, $T_{\mathrm{step}}:= [T^1_{\mathrm{step}}, \dots, T^{N_{\mathrm{meas}}}_{\mathrm{step}}]$, $\Phi_{\mathrm{cont}}:= [\Phi^1_{\mathrm{cont}}, \dots, \Phi^{N_{t}}_{\mathrm{cont}}]$.}
\end{figure*}

Fig. \ref{psfs} shows $\Phi_{\mathrm{step}}$ and $\Phi_{\mathrm{cont}}$ in reflection and transmission configuration, respectively. To reduce the huge amount of data generated by performing step scanning measurements, we applied the maximum thermogram (MT) method \cite{paper1} to the step scanning data which eliminates the time dimension from $T_{\mathrm{meas,step}}\in \mathbb{R}^{N_r\times N_t\times N_\mathrm{meas}}$ by selecting the time stamp $t=t_{\mathrm{MT}}$ where the maximum temperature amplitude is reached (c.f. Fig. \ref{flow chart}). Effectively, this leads to $T_{\mathrm{step}}\in \mathbb{R}^{N_r\times N_\mathrm{meas}}$ and a vertical section through $\Phi_{\mathrm{step}}:=\Phi_{\mathrm{step}}(r,z=0|L,t=t_{\mathrm{MT}})$ in Fig.\ref{psfs} a,b), represented by the dashed vertical lines. Consequently, we reformulate equation (\ref{main_eq}) and describe the measured data for step and continuous scanning by the following two equations:
\begin{ceqn}
\begin{equation}
    T^i_{\mathrm{step}}[k] = \big(\Phi_{\mathrm{step}} * x_{\mathrm{step}}^i\big)[k] ~~~~~~~~~~~~~~~~~~~~~~~~~~~~~~
    T^i_{\mathrm{cont}}[k] = \big(\Phi_{\mathrm{cont}}^i * x_{\mathrm{cont}}^i\big)[k].
\end{equation}\end{ceqn}
These are sets of 1dim problems in the $r$-domain.
Since we are measuring the data, we have to describe the measured data by discrete values $T^i_{\mathrm{step/cont}}[k] = T^i_{\mathrm{step/cont}}(k \cdot \Delta r_{\mathrm{cam}})$ with $k = 1 \dots N_r$. 
It should be noted that the measured continuous scanning data indeed considers the time, but here we interpret each time stamp as a measurement $i$. 



\subsection*{Inspection of internal defects using IJOSP}

The iterative joint sparsity (IJOSP) approach promotes joint sparse solutions and is particularly of interest if blind illumination is used \cite{murray2017super,haltmeier2013block}. In this work, the data is treated as if we had measured blindly, since we consider the position of the illumination in $x$ and not in $\Phi$. This is a worst-case scenario, which is of practical nature, because we do not always know the exact position of the illumination. Therefore, the IJOSP approach empirically enables us to find a solution $\hat{x} \approx x$ and to describe the defect pattern in space considering the following minimization problem: 
\begin{ceqn} \begin{equation} \label{ijosp}
    \min_{\hat{x}} \frac{1}{2} \sum_{i=1}^{N} \sum_{k=1}^{N_r}\big| \big(\Phi^i \ast \hat{x}^i\big)[k] - T^i[k]\big|^2 + \lambda_1 \|\hat{x}\|_{2,1} +  \frac{\lambda_2}{2} \|\hat{x}\|_2^2,
\end{equation}\end{ceqn}
for step scanning ($\Phi^i=\Phi_{\mathrm{step}}, T^i=T_{\mathrm{step}}^i, \hat{x}^i=\hat{x}_{\mathrm{IJOSP,\,step}}^i,N=N_{\mathrm{meas}}$) and continuous scanning ($\Phi^i=\Phi_{\mathrm{cont}}^i, T^i=T_{\mathrm{cont}}^i, \hat{x}^i=\hat{x}_{\mathrm{IJOSP,\,cont}}^i,N=N_t$), respectively. $\lambda_1$ and $\lambda_2$ are controlling the regularizers, more precisely, the impact of block sparsity $\|\hat{x}\|_{2,1} = \sum_{k=1}^{N_r} \sqrt{\sum_{i=1}^{N} |\hat{x}^i[k]|^2}$ and Tikhonov regularization $\|\hat{x}\|_2^2$. 

To find a good solution with $\hat{x} \approx x$, there are several ways to solve the minimization problem in equation (\ref{ijosp}). In this work we have used the optimization algorithm Block-FISTA (does not make use of Tikhonov regularization) and Block-Elastic-Net method \cite{ahmadi2,fista,zou}. Both optimization methods use an updating step size related factor $L_c$ within the gradient descent implementation which is richly described in literature as the Lipschitz constant for the gradient of the least squares error term shown in equation (\ref{ijosp}). The number of updates is described by the number of iterations $N_{\mathrm{iter}}$.

\subsection*{Quantitative evaluation of the final reconstruction result}
The final reconstruction result $\hat{x}_{1D}$ as shown in the flow chart in Fig. \ref{flow chart} is calculated by the sum over all measurements of $\hat{x}$. Since SR experiments are performed with multiple measurements and small position shifts resulting in overlaps between the measurements, the normalized sum of all illuminations $I^i$ over the number of measurements equals an one-array over all pixels. This means that the sum over all measurements of $\hat{x}$ would approximately equal to $a$. In the following we work with $a_{\text{rec}} = \hat{x}_{1D}/\max\{\hat{x}_{1D}\} = \sum_i \hat{x}^i/\max\{\sum_i \hat{x}^i\}$ as final result to quantitatively evaluate and compare with the originally manufactured $a$ of the specimen.

\begin{figure*}[h]
\centering
\includegraphics[width=0.8\textwidth,trim=2cm 24.5cm 4cm 0cm, clip]{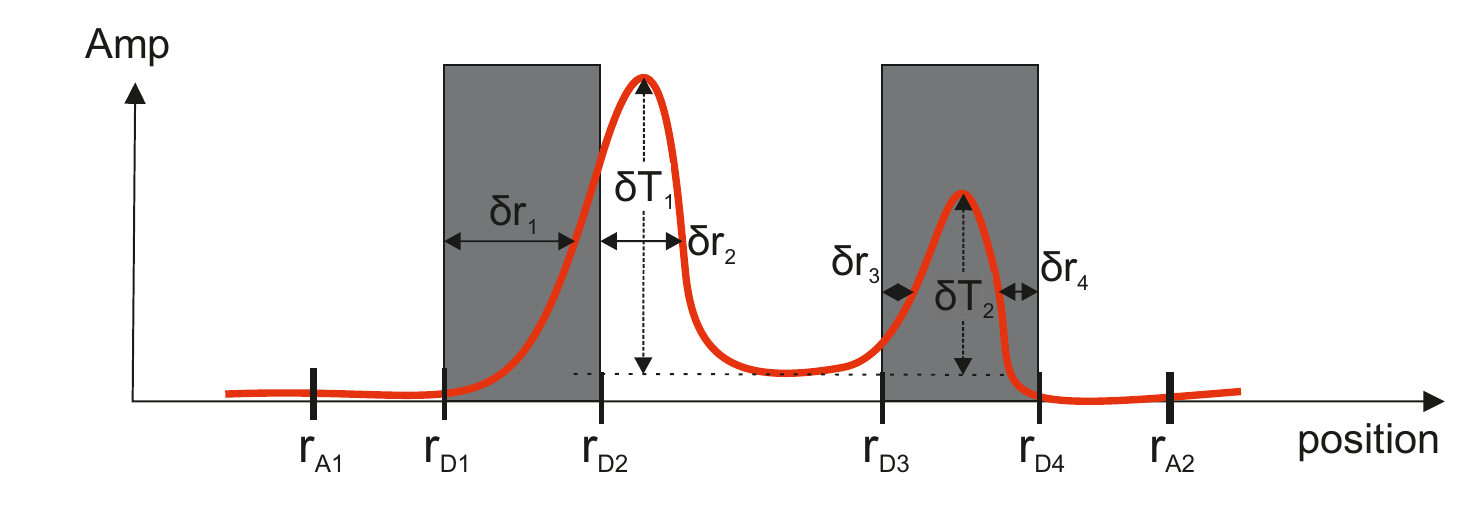}
\caption{\label{qnde} Explanation of metrics for QNDE: $\delta T$ and $\delta r$. The red curve illustrates an exemplary $a_{\text{rec}}$ and the grey defect pattern shows the corresponding exemplary $a$.}
\vspace{-2mm}
\end{figure*}

For quantitative nondestructive evaluation (QNDE), the following metrics are used: $\delta T$ and $\delta r$. Fig. \ref{qnde} explains these metrics.

Thus, $\delta T$ represents the contrast between defective and non-defective area investigating one specific defect pair. Further, $\delta r$ denotes the distance between one peak in $a_{\text{rec}}$ to the real defect position in $a$ referring to the full width half maximum (FWHM). Moreover, $\delta T_{1/2} \in [0,\delta T_{\text{1/2,\,max}}$] with $\delta T_{\text{1/2,\,max}} = 1$ due to the implemented normalization and $\delta r_{1/2/3/4} \in [0,\delta r_{1/2/3/4,\,\text{max}}$]. Assuming that $r_{A1}$ and $r_{A2}$ limit the investigated area according to Fig. \ref{qnde}, the maximum values can be calculated by: $\delta r_{1,\,\text{max}} = r_{A2}-r_{D1}$, $\delta r_{2,\,\text{max}} = r_{A2}-r_{D2}$, $\delta r_{3,\,\text{max}} = r_{D3}-r_{A1}$ and $\delta r_{4,\,\text{max}} = r_{D4}-r_{A1}$. The reconstruction accuracies (ra) for $\delta T$ and $\delta r$ can then be given by:
\begin{ceqn}
\begin{equation} \label{eq:ra}
    \text{ra}_{\delta T} = \frac{1}{2}(\delta T_1 + \delta T_2), ~~~~~~~~~~~~
    \text{ra}_{\delta r} = \frac{1}{4}\bigg(\frac{\delta r_{1,\,\text{max}}-\delta r_1}{\delta r_{1,\,\text{max}}}+\frac{\delta r_{2,\,\text{max}}-\delta r_2}{\delta r_{2,\,\text{max}}}+\frac{\delta r_{3,\,\text{max}}-\delta r_3}{\delta r_{3,\,\text{max}}}+\frac{\delta r_{4,\,\text{max}}-\delta r_4}{\delta r_{4,\,\text{max}}}\bigg).
\end{equation}\end{ceqn}
The overall reconstruction accuracy for an area $A$ covering one defect pair is further determined by: $\text{ra}_A = 0.5 \cdot (\text{ra}_{\delta T}+\text{ra}_{\delta r})$.

\section*{Results and discussion}
\begin{figure*}[h]
\centering
\includegraphics[width = 0.7 \textwidth, trim = {45cm 10cm 20cm 0cm},clip]{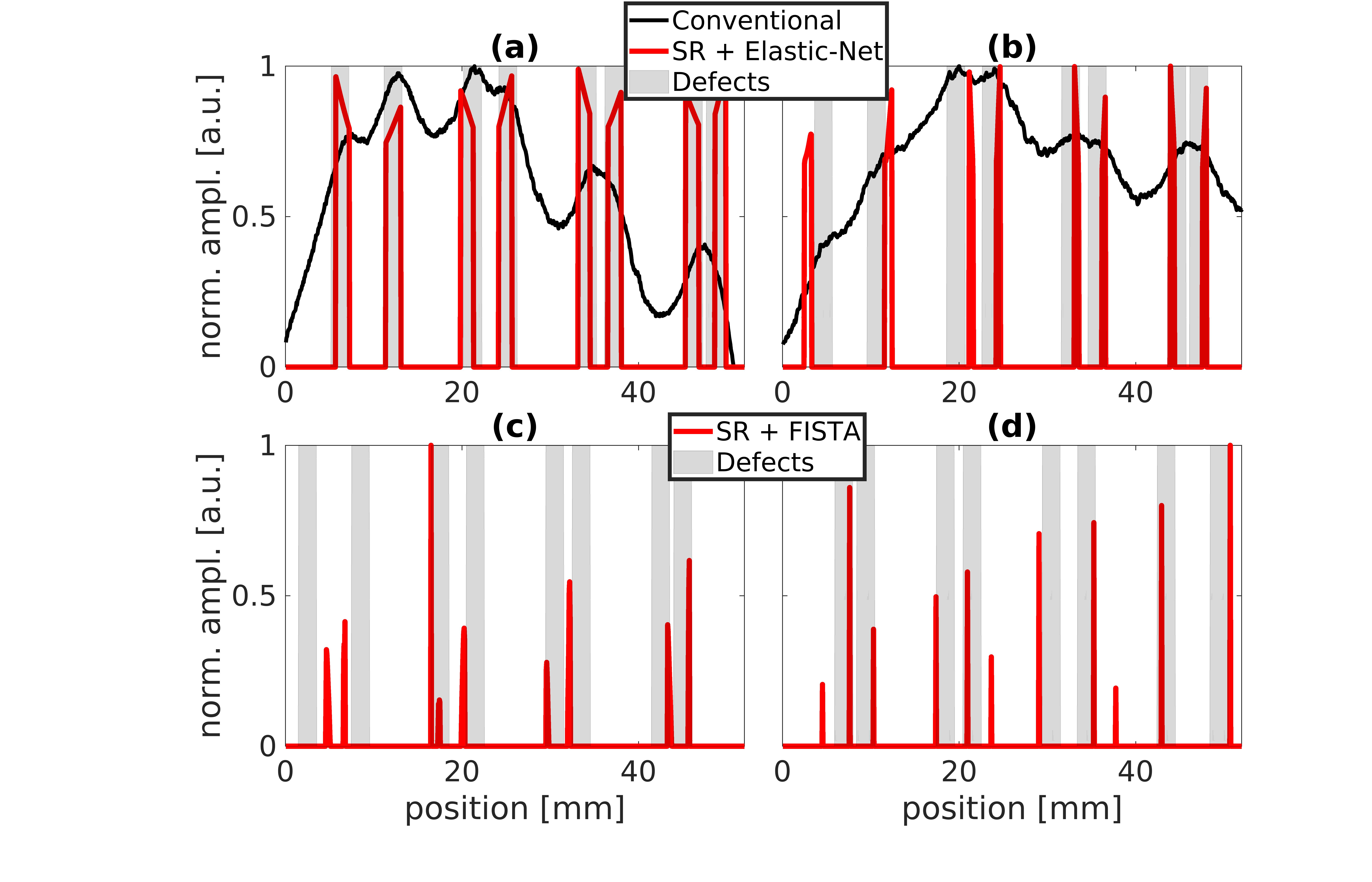}
\caption{\label{results_ijosp} Normalized IJOSP results: a) step scanning + refl.; b) step scanning + trans.; c) continuous scanning + refl.; d) continuous scanning + trans. The diagrams in a) and b) show additionally the conventional result obtained by simply applying the maximum thermogram method to the raw data of a measurement where the whole sample surface is illuminated - here we illuminated the whole sample surface with a laser square $60 \times 60\,\text{mm}^2$, $t_{\mathrm{pulse}}\,=\,2$\,s.}
\vspace{-2mm}
\end{figure*}

Fig. \ref{flow chart} shows the processing steps from the raw data averaged along $y$ to the final 1dim reconstruction results of the investigated internal defects shown in Fig. \ref{results_ijosp}. To obtain these final results (red curves), we have used the parameters listed in Table \ref{table_parameters}.

\begin{table}[!h]
\centering
\begin{tabular}{|c||c|c|c|c|}
\hline
\textbf{parameters} & \textbf{(a)} & \textbf{(b)} & \textbf{(c)} & \textbf{(d)}\\
\hline
\hline
position shift $\Delta r$ [mm] & 0.2 & 0.2 & - & -\\
\hline
$t_{\mathrm{pulse}}$ [ms] & 500 & 500 & - & - \\
\hline
$v_{\mathrm{scan}}$ [mmps] & - & - & 1 & 50 \\
\hline
measurement duration [min] & 30 & 30 & 1 & 1/30 \\
\hline
\hline
Lipschitz const. $L_c$ [a.u.] & 1.41 & 4.24 & 7.07 & 4.95 \\
\hline
$\lambda_1$ [a.u.] & 2.38 & 20 & 50 & 14 \\
\hline
$\lambda_2$ [a.u.] & 0.05 & 0.0005 & 0 & 0 \\
\hline
$N_{\mathrm{iter}}$ [a.u.] & 500 & 500 & 500 & 2500 \\
\hline
computation time [s] & 41.67 & 57.06 & 132.23 & 18.8 \\
\hline
\end{tabular}
\caption{\label{table_parameters} Chosen experimental and processing parameters to obtain the IJOSP results (red curves) in Fig. \ref{results_ijosp}}
\end{table}

 Analyzing Fig. \ref{results_ijosp} clearly shows that we are able to resolve all internal defects in reflection and transmission configuration for the additively manufactured stainless steel specimen. Especially in the step scanning reflection case (see Fig. \ref{results_ijosp} (a)) we obtain outstanding results being able to reconstruct each internal defect almost perfectly. Even with continuous scanning (Fig. \ref{results_ijosp} (c,d)) we are able to resolve the internal defects, but not as accurate as in step scanning. Here, the most prominent peaks indicate the correct positions of the investigated internal defects, but sometimes some minor peaks occur as small artifacts as well. But of course we also have to take into account that we have to measure two to three orders of magnitude longer for the step scan (c.f. Table \ref{table_parameters}). Furthermore, it is noticeable that the red result curves in diagrams (a,b) exhibit broader peaks than the narrow peaks in (c,d). This comes from the fact that in (a,b) we have used the Block-Elastic-Net optimization within IJOSP approach that makes use of the Tikhonov regularization ($\lambda_2 > 0$, c.f. Table \ref{table_parameters}) which smoothens the signal in the spatial sparsity domain, whereas the Block-FISTA ($\lambda_2 = 0$) algorithm is used in (c,d). 
 
 The results can be further examined quantitatively by using the reconstruction accuracy as defined in the eq. \ref{eq:ra}. Table \ref{recaccuracies} shows all calculated values based on the resolution accuracy given by the specification of the used IR camera in terms of spatial resolution $\Delta r_{\text{cam}} = 54\,\mu$m and temperature resolution $\Delta T_{\text{cam}} = 25\,$mK.
 
 \begin{table}[h]
\centering
\begin{tabular}{|c||c|c|c|c|c|c||c|}
\hline
 & \textbf{(a) conv.} & \textbf{(a) SR} & \textbf{(b) conv.} & \textbf{(b) SR} & \textbf{(c) SR} & \textbf{(d) SR} & \textbf{defect geometries} (distance:width:depth)\\
\hline
\hline
$\text{ra}_{A1}$ & 0.52 & 0.92 & 0.5 & 0.8 & 0.58 & 0.88 & $4:2:2\,$mm \\
\hline
$\text{ra}_{A2}$ & 0.5 & 0.91 & 0.52 & 0.84 & 0.78 & 0.79 & $2:2:2\,$mm  \\
\hline
$\text{ra}_{A3}$ & - & 0.93 & 0.47 & 0.86 & 0.63 & 0.71 & $1:2:2\,$mm \\
\hline
$\text{ra}_{A4}$ & - & 0.92 & - & 0.85 & 0.69 & 0.76 & $0.5:2:2\,$mm \\
\hline
$\text{ra}$ & 0.26 & 0.92 & 0.37 & 0.84 & 0.67 & 0.79 & \cellcolor{black!25}\\
\hline
\end{tabular}
\caption{\label{recaccuracies}Calculated reconstruction accuracies $\text{ra} \in [0,1]$ with $\text{ra} = 0.25\cdot (\text{ra}_{A1}+\text{ra}_{A2}+\text{ra}_{A3}+\text{ra}_{A4})$ for each shown result in Fig. \ref{results_ijosp}, $A1$ indicates the area of the defect pair with the largest distance to each other and $A4$ the defect pair area with the shortest distance to each other. Table cells with a dash inside mean that it was not possible to identify two defects. The defect geometries in the last column are speciefied by the distance between two defects within a defect pair, the width of one defect (all defects have the same width) and the depth of the defects (all defects have the same depth). A graphical visualization of the investigated defect geometries is shown in Fig. \ref{exp_setup} b).}
\end{table}
Table \ref{recaccuracies} confirms the above mentioned statements based on the qualitative analysis. In addition, it can be seen that the use of SR results in reconstruction accuracies of over 90\,\% whereas conventional methods fail completely (see $\text{ra}_{A3}$ and $\text{ra}_{A4}$). Even the reconstruction based on continuous scanning data, that is of particular interest for industrial applications, exhibits reconstruction accuracies of around $70-80\,$\%.
Consequently, the quantitative analysis shows that very high reconstruction accuracies ($>90\,\%$) can be achieved with the proposed method only for laser step scan measurements. If an exact reconstruction is not necessary, but only defects should be detected, the continuous scanning could be sufficient as data basis, which has the advantage that these data can be generated much faster than with the step scan.


\section*{Conclusion and outlook}

In this study we could show that we are able to resolve closely spaced internal defects inside an additively manufactured stainless steel specimen. We obtained accurate results in 1D reconstruction outperforming conventional thermographic methods realized by homogeneous illumination of the whole sample surface. The resulting diagrams after applying IJOSP show an at least four times better spatial resolution since we could easily separate defects with a defect depth to defect distance ratio up to 4:1 instead of 1:1 in conventional thermographic inspection techniques. The best results have been obtained in step scan reflection mode but the very fast continuous scanning mode does lead to convincing results as well. To obtain these outstanding results, we have used standard measurement technology, as found in many thermography laboratories, but upgraded with a SR measurement strategy and a CS-based post-processing. Thus, these studies encourage the use of SR laser thermography in metal industry for an accurate inspection of e.g. production samples with blowholes or other inclusions.

However, the regularization parameters within the IJOSP approach have been chosen manually. Therefore, as an outlook, we are working on a deep neural network approach to figure out the optimal regularization parameter based on simulated and/ or experimental training data.

\section*{Data availability statement}
The data that support the findings of this study are available from the corresponding author
upon reasonable request.

\bibliography{sample}


\section*{Acknowledgements}
The work of P. Burgholzer was supported by the Austrian Science Fund (FWF), projects P 30747-N32 and P 33019-N.

\section*{Author contributions statement}

S.A., P.B. and M.Z. conceived the idea of super resolution laser thermography; S.A., P.B., P.J. and G.C. contributed to the compressed sensing based signal processing; S.A., P.D.H. and M.Z. designed the specimen; S.A. and J.L. conducted the experiments; S.A. wrote the manuscript with support from M.Z.; P.D.H. and J.L. contributed to the preparation of figure 1, 2 and 4. All authors reviewed the manuscript. 

\section*{Additional information}

\textbf{Competing interests}: The authors declare no competing interests.

\end{document}